\documentclass[pdflatex,sn-nature]{sn-jnl}

\usepackage{graphicx}%
\usepackage{multirow}%
\usepackage{amsmath,amssymb,amsfonts}%
\usepackage{amsthm}%
\usepackage{mathrsfs}%
\usepackage[title]{appendix}%
\usepackage{xcolor}%
\usepackage{textcomp}%
\usepackage{manyfoot}%
\usepackage{booktabs}%
\usepackage{algorithm}%
\usepackage{algorithmicx}%
\usepackage{algpseudocode}%
\usepackage{listings}%

\theoremstyle{thmstyleone}%

\theoremstyle{thmstyletwo}%

\theoremstyle{thmstylethree}%

\newcommand{\ie}{\textit{i.e.}, }
\newcommand{\eg}{\textit{e.g.}, }

\begin{document}

\title[Structure recovery from atomistic descriptors]{FUCrIMODo: structure recovery from atomistic descriptors via multi-stage genetic algorithms}


\author*[1]{\fnm{Louis} \sur{B\"ohm}}\email{louis.boehm@gmx.de}

\author[1]{\fnm{Martin} \sur{Kuban}}

\author[1]{\fnm{Claudia} \sur{Draxl}}

\affil[1]{\orgdiv{Department of Physics and CSMB}, \orgname{Humboldt-Universität zu Berlin}, \orgaddress{\street{Zum Gro\ss en Windkanal 2}, \city{Berlin}, \postcode{12489}, \state{Berlin}, \country{Germany}}}

\abstract{Data-driven approaches to materials discovery rely on numerical representations of atomic structures as input for machine learning models. Inverting these descriptors — recovering atomic structures from their representations — is essential for most generative material design pipelines, yet it remains challenging, particularly for periodic systems. Existing inversion methods are either tailored to specific invertible descriptors or require candidate structures with similar atomic arrangements and compositions, limiting the exploration of novel regions in chemical and configurational space. Here, we propose a generalizable, similarity-driven sampling approach, powered by a novel stage-wise optimization strategy, to recover atom types, atomic positions, and unit cell shapes directly from a descriptor. Our approach requires only descriptor features and parameters as input without any prior structural knowledge. The capability of our method is demonstrated by the averaged Smooth Overlap of Atomic Positions (SOAP) descriptor.}




\maketitle

\section{Introduction}\label{sec:Introduction}

In recent years, data driven approaches to materials discovery and development have become an integral component of our research landscape~\cite{Tanaka2018, Data_Driven_Mat_Himane_2019, Accelerating_ma_Pyzer_2022, Generative_AI_f_De_Bre_2025}, where enormous amounts of data are used to train statistical models \cite{Foundation_mode_Pyzer_2025}. These models can be used to predict the structure and properties of existing or hypothetical materials, which, in turn, make screening of large numbers of candidates feasible~\cite{Schmidt2023, Cerqueira2023, Wines2024}. Given the popularity of this approach, many research groups are training and publishing machine-learning (ML) models for different purposes~\cite{Schmidt2019}. However, these models are rarely re-used in follow-up projects. An important reason for that is that often, these models perform best for the data they have been trained on~\cite{Schmidt2023, Speckhard2025}. Therefore, new applications often require to re-train, or at least fine-tune, a model for a specific task. At least equally important is the fact that models typically can't be inverted. In other words, while the quality of a model can be validated by cross validation (CV) on test data, there is no practical way of finding new materials that satisfy it rather than by further sampling, which requires additional resources. To make data-based research more sustainable, one should find ways to make the most of existing models.

Why is model inversion challenging in the materials-science domain? A key reason is that materials must first be represented by descriptors before training ML models can start. Descriptors encode the selected properties of the material~\cite{Big_Data_of_Mat_Ghirin_2015, Ward2016} or its atomic structure~\cite{On_representing_Bartok_2013, Neural_Message_Gilmer_2017, Crystal_Graph_C_Xie_T_2018}. During the training process, the model establishes a mapping between these descriptors and the property that the model should predict. The complexity of this mapping depends heavily on the choice of the descriptor: For example, if the descriptor does not encode the symmetries of the material, the ML model must learn them explicitly~\cite{DScribe_Librar_Himane_2020}. Therefore, many successful descriptors account for such symmetries~\cite{Valle2010, On_representing_Bartok_2013, Big_Data_of_Mat_Ghirin_2015, DScribe_Librar_Himane_2020}. Unfortunately, for the same reason, such descriptors typically can not be inverted, since in the general case, the unit cell and the atom positions cannot be recovered from the descriptor.

We note that we explicitly distinguish between model inversion and \textit{generative} models~\cite{Generative_AI_f_De_Bre_2025}. The latter are trained specifically for the purpose of generating novel crystal geometries with specific properties. These models directly establish a mapping between the crystal structure and the targeted property. This mapping can be sampled under a specific condition, \ie a desired property of the generated structure~\cite{Generative_AI_f_De_Bre_2025, A_generative_mo_Zeni_2025, DeBreuck2026}. Such models are often based on neural-network (NN) architectures and are trained for high-throughput (HT) purposes and need large amounts of training data. Training data for material properties are, however, very scarce. Furthermore, generative models also rely on an invertible description of the crystal structure, which is typically achieved by maintaining a real-space representation of the compound alongside with the descriptor\cite{A_generative_ma_Breuck_2025}.

A typical way of inverting ML models is using a sampling algorithm or a generative ML model for generating atomic structures and predicting the properties of these candidates~\cite{Generative_AI_f_De_Bre_2025}. The sampling is continued until one or several candidates with the desired properties are found. A downside of this approach is that it depends on the ability of the sampling algorithm to generate meaningful new structures. An alternative way is to sample descriptors that, according to a ML model, correspond to materials with desired properties, and finally invert the descriptor to retrieve the corresponding crystal structure. This approach has been demonstrated with a specialized, invertible descriptor for periodic systems to predict 14 novel narrow-gap semiconductors~\cite{An_invertible_Xiao_2023}. The descriptor used in that work represents atomic structures as strings and is constructed from graph representations of crystal structures. From this descriptor, the atomic structure was obtained by applying a series of transformations, including the generation of structures and their relaxation using two different force fields. With this approach, it was possible to retrieve 84.66 \% of structures consisting of up to 20 atoms in a benchmark test~\cite{An_invertible_Xiao_2023}.

In this work, we propose a similarity-driven sampling approach based on a genetic algorithm (GA) to recover atom types, atomic positions, and unit-cell shapes from arbitrary structural descriptors. This method requires only descriptor features and the parameters that were used for calculating them. In other words, no prior knowledge of the structural configuration is required. We demonstrate it's applicability on an averaged Smooth Overlap of Atomic Positions (SOAP) descriptor~\cite{On_representing_Bartok_2013}, significantly surpassing previous approaches, which were based on gradient descent \cite{Local_inversion_Cobell_2022, On_representing_Bartok_2013}.

\section{Results}\label{sec:Results}

\subsection{Descriptor inversion through effective sampling}

In the context of this work, a \textit{descriptor} is a numerical vector representing the atomic structure of a material. The \textit{target} descriptor is the descriptor associated to an unknown compound of interest, \eg one that was suggested by a ML model. To find the atomic structure of the target descriptor, we compare it to the descriptors of \textit{candidate} structures, which we obtain through sampling. The comparison is done using a suitable similarity metric $S$. If the similarity is sufficiently high, \ie exceeding a predefined threshold $S_{thres}$, the candidate structure is accepted as the inversion of the target descriptor. Efficiently sampling the space of candidate structures is central to our approach, yet several factors make this task particularly challenging. First, the pool of possible structures is infinitely large. This is in part due to the invariance of most structural descriptors with respect to operations such as rotations and translations of the input cell as well as the permutation of atoms \cite{DScribe_Librar_Himane_2020}. Furthermore, most \textit{global} descriptors, \ie those that describe the full atomic structure of a compound, are intensive, \ie the descriptor for a unit cell and any supercell of it are identical. Beyond these fundamental challenges, the computational cost of computing the descriptors can be significant, since they are generated for each candidate structure during sampling. For descriptors that are based on atomic environments, such as SOAP, the computational cost further increases with the size of the structure. Moreover, when using a sampling algorithm, the threshold for the magnitude of changes to the structure imposed by the algorithm must be carefully chosen: Too small changes lead to slow convergence, while too large changes increase the risk of missing the global minimum of the optimization. We address these challenges by using a specialized GA. Inspired by biological principles, GAs work by optimizing a \textit{population} -- here, candidate structures -- towards a specific goal. The goal is encoded in a \textit{fitness function} -- here, among other criteria, the similarity to the target descriptor. The optimization is performed iteratively, \ie a new population is formed in each step by manipulating the candidate structures and assessing their performance based on the fitness function. In an extension of this well established algorithm, we introduce stage-wise optimization, \ie during the runtime of the GA, we define \textit{stages}, which are characterized by a specific selection of GA parameters. Each stage is designed to reach a specific optimization task, such as increasing the diversity of candidate structures or optimizing towards local minima of the fitness function. Multiple stages can then composed, in sequence or in parallel, to form the complete optimization workflow.

\subsection{Algorithm}

\begin{figure}
	\centering
	\includegraphics[width=\textwidth]{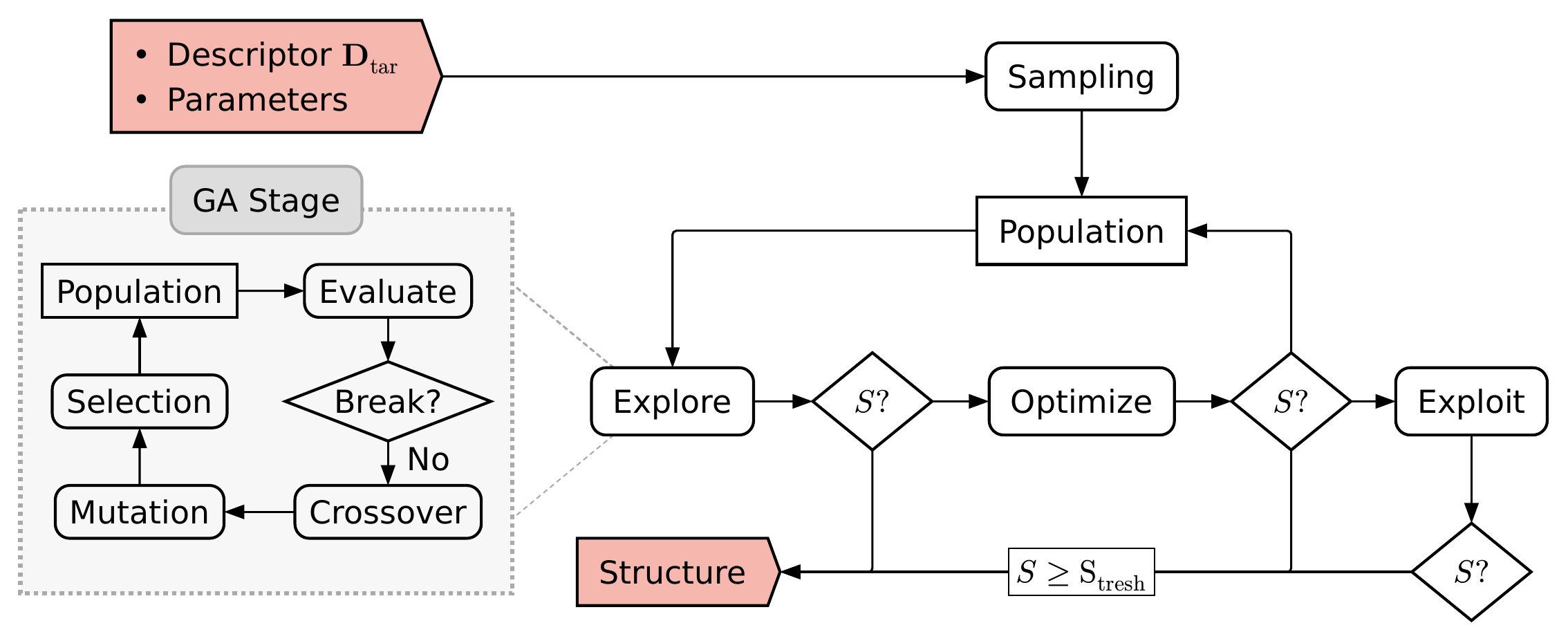}
	\caption{Workflow for finding crystal structures from global SOAP descriptors. The input to this workflow are a target descriptor \textbf{D}$_{tar}$ and the parameters used to generate it. Then, a population of initial structures is generated through \textit{sampling}. The \textit{population} is then optimized with consecutive GA stages, \textit{explore}. \textit{optimize}, and \textit{exploit} (see text for details). If, at any stage of the algorithm, a structure with a sufficiently high similarity $S \geq$ S$_{\mathrm{tresh}}$ to \textbf{D}$_{tar}$ is found, the algorithm terminates and the structure is returned as the result. The individual steps of a GA stage (gray box) are \textit{evaluating} the population, the subsequent application of \textit{crossovers} and \textit{mutations}, and the \textit{selection} of high performing structures to form the new \textit{population}. Red boxes mark in- and output, squared boxes represent the population of structures, rounded boxes show processing steps applied to the population, and diamond shaped ones display decision points. }
	\label{fig:workflow-overview}
\end{figure}

Figure~\ref{fig:workflow-overview} illustrates the general workflow of our approach. Methodological details for all parts of the algorithm can be found in Section~\ref{sec:Methods}. In the first step, structures are generated by random sampling to act as an initial population. The similarity $S$ between the target descriptor \textbf{D}$_{tar}$ and the descriptor $\mathbf{D}$(C) of a candidate structure C is calculated. Typically, the similarity of any initial random structure to the target is very close to 0. After this initialization, the algorithm enters its iterative phase. In each generation, members of the population are selected based on their fitness and manipulated by two mechanisms: mutations and crossovers. Mutations are distortions of the candidate, \eg random displacements of atoms in the structure. Cross-overs are operations that use two members of the population and combine their properties, \eg the exchange of atom types between structures. During an iteration, the algorithm may perform either one of these manipulations, both, or none. These operations create new candidates, which are added to the population. (Tables with all possible mutations and crossovers can be found in the Supporting Information.) In a next step, the population size is reduced by applying a structure-selection algorithm.

Which and how many candidate structures are generated, is controlled by different parameters of the algorithm. For each stage of the algorithm, different GA parameters are chosen to control its overall convergence behavior. We distinguish between three types of stages: During \textit{exploration}, the algorithm aims to scan unvisited regions in the configurational space by generating structures with entirely new atomic configurations and unit-cell shapes. In the \textit{exploitation} stage, existing structures are slightly adjusted to quickly find nearby extrema in the search space. During \textit{optimization}, the goal is to improve found structures while simultaneously keeping diversity high. During the execution of the algorithm, the different types of stages are executed sequentially, where the first stage is an exploration stage, the second is an optimization stage, the third is an exploitation stage. This sequence marks one iteration of the inversion algorithm and repeats until the breaking condition is reached in any of the stages, \ie a structure is found with a similarity to the target descriptor above the predefined threshold $S \geq$ S$_{\mathrm{tresh}}$, or the maximal number of stages is exhausted.

\subsection{Examples and analysis}

In the following, we illustrate the stages of the optimization and the structure-inversion process using three known compounds as examples. They are selected from the MP-20 dataset~\cite{MP20-Github-Zugner}, a subset of the Materials Project database~\cite{Commentary_The_Jain_2013}. Their structures are used to compute global SOAP descriptors~\cite{On_representing_Bartok_2013} as targets $\mathbf{D}_\mathrm{tar}$ for the inversion algorithm. Details on the descriptor generation can be found in Section~\ref{sec:Methods}. The original structure is not used in any step of the algorithm, but only for validation of our results.

\begin{figure}
	\centering
	\includegraphics[width=0.86\textwidth]{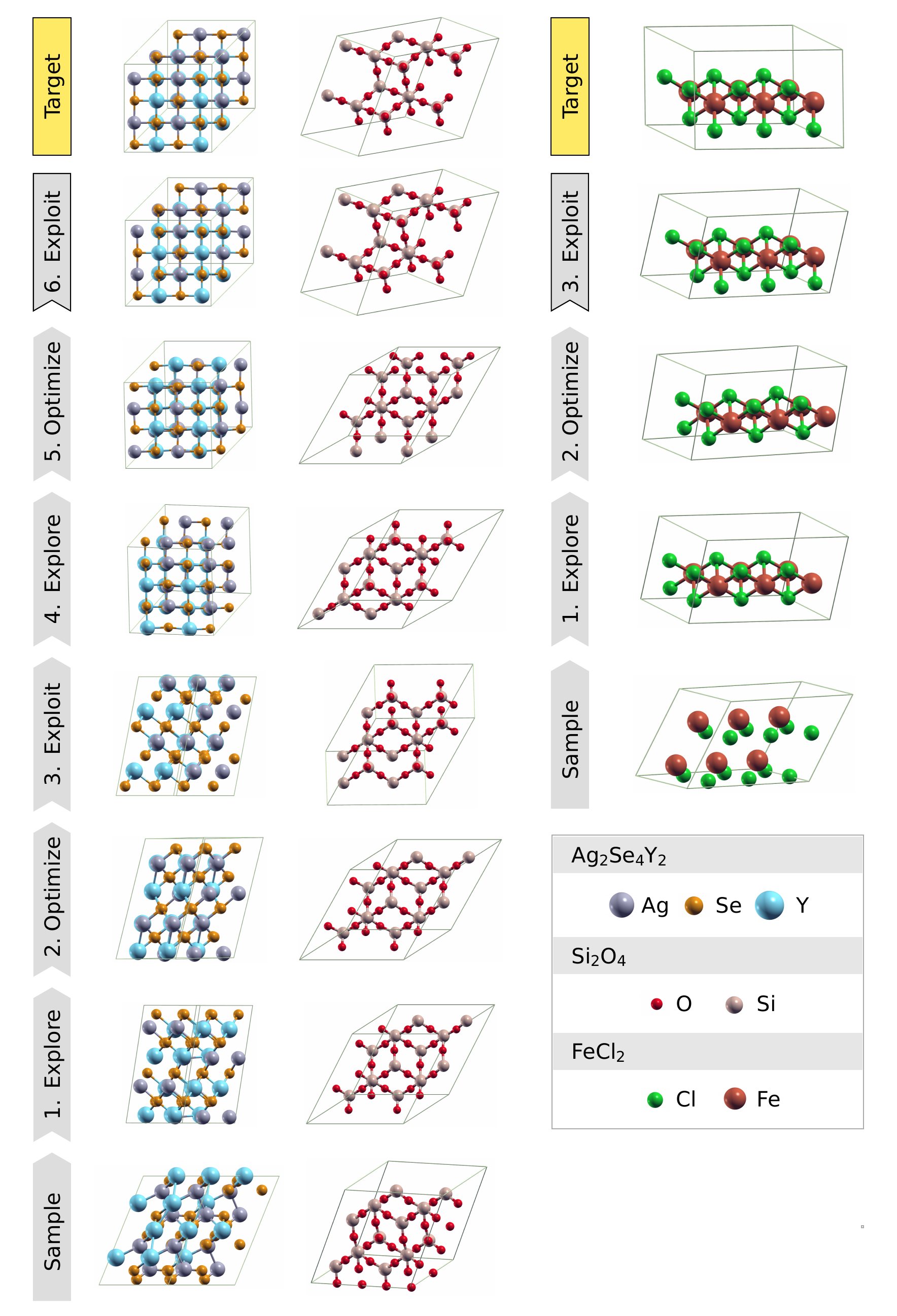}
	\caption{Structural evolution during the inversion of the global SOAP descriptor demonstrated by three examples. Within a column, the atomic structures shown from bottom to top, illustrate how the algorithm progressively reconstructs the crystal structure that reproduces the target descriptor. The target structures, used to calculate the target descriptor, are shown in the top row. Structures are displayed as supercells for better visual comparability. Ag$_2$Se$_4$Y$_2$ (left column), Si$_2$O$_4$ (middle column) require six stages to be found, while FeCl$_2$ (right column) is identified within only three stages.}
	\label{fig:structure-development}
\end{figure}

The structural evolutions of the descriptor inversion process are illustrated in Fig. \ref{fig:structure-development}, going from bottom to top.
The top row displays the target structures. The first example (left), Ag$_2$Se$_4$Y$_2$, has a triclinic structure in which Ag and Y occupy octahedral sites around a regular Se sublattice. The second example (center), Si$_2$O$_4$, has a hexagonal unit cell. Each Si atom is tetrahedrally coordinated by four O atoms. These tetrahedra are linked, as every O atom has two Si neighbors. The third example (right), FeCl$_2$, is a two-dimensional monolayer with  hexagonal symmetry, where Fe atoms are situated between the Cl atoms at the surface. The bottom panels show the structures of the initial population, obtained by random sampling, the higher panels show the best performing candidates in each stage, generated by our algorithm.

Let us focus first on Ag$_2$Se$_4$Y$_2$, the example shown on the left. The structure obtained by random sampling differs significantly from the target structure. The unit-cell vectors are different, and the arrangement of atoms in the unit cell appears less ordered. The structures generated in the first two stages are almost monoclinic (two angles different from 90$^{\circ}$, although close). In both structures, Ag and Y atoms are positioned between Se atoms showing an ordered atomic arrangement similar to the target structure. In the third stage, the generated structure remains monoclinic, with atoms in a well-ordered arrangement. In the later stages of the algorithm, the structures become visibly more similar to the target structure, and are basically indistinguishable in stage 6.

In the second example, shown in the center, the sampled structure shows already a similar atomic coordination to that of the target structure, however, the shape of the unit cell is different. The sampled structures, especially their unit-cell shapes, become more similar to the target structure during the runtime of the algorithm. The coordination is kept intact at all stages. In the last example, FeCl$_2$, the random sampling generated two individual FeCl monolayers as starting configuration. Interestingly, the algorithm requires only three stages to reconstruct the target structure.

\begin{figure}
	\centering
	\includegraphics[width=\textwidth]{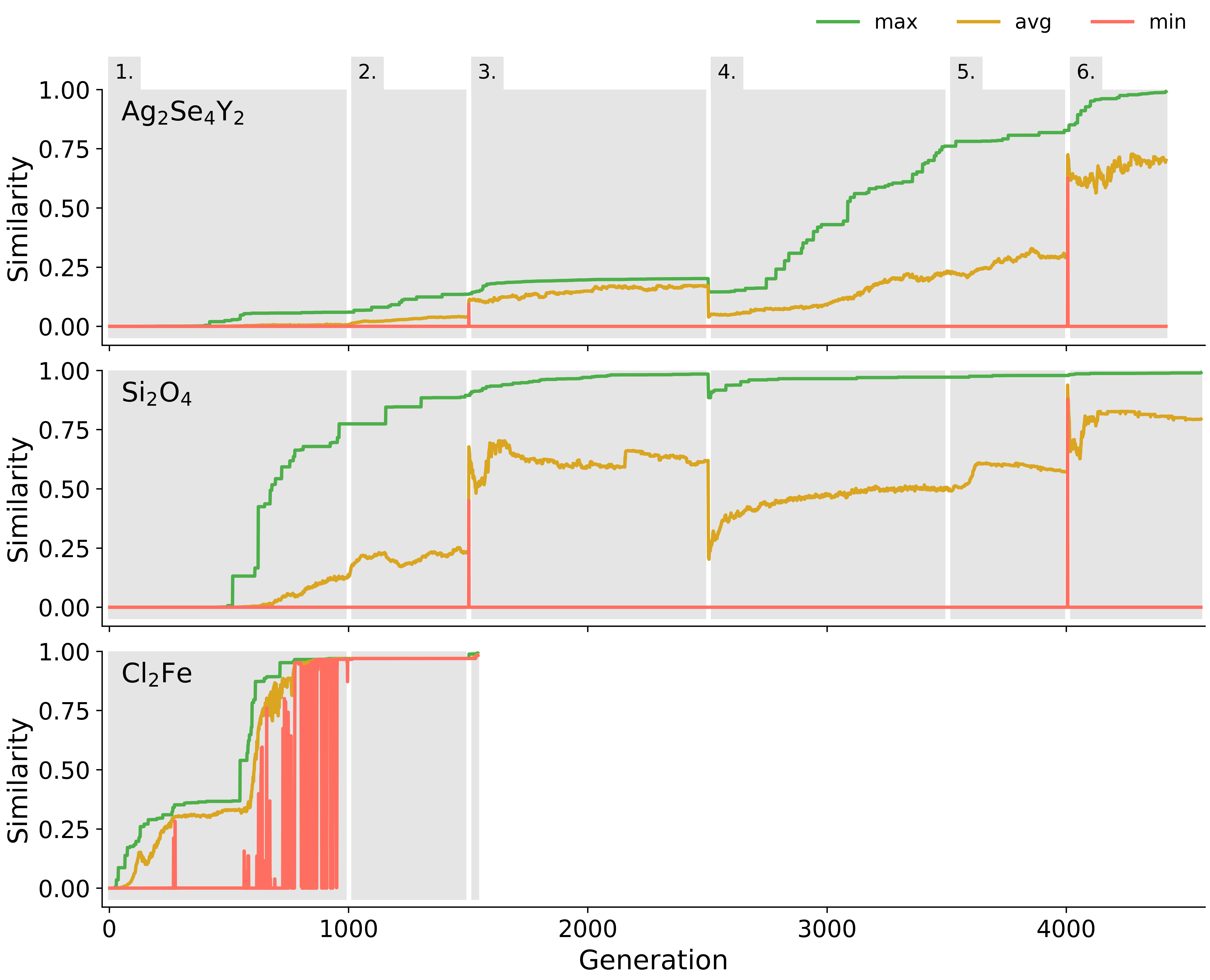}
	\caption{Convergence of the descriptor inversion algorithm for descriptors calculated from Ag$_2$Se$_4$Y$_2$ (top), Si$_2$O$_4$ (center), and FeCl$_2$ (bottom). Values of minimum (red), maximum (green), and average (yellow) similarity S$_{0.1}$ of all candidate structures within the population are shown for each GA generation. Gray shaded regions mark the six consecutive stages of the algorithm.}
	\label{fig:similarity-development}
\end{figure}

The evolution of the structures can also be described from a more statistical standpoint. To do so, Fig.~\ref{fig:similarity-development} shows for each of the three examples the minimum, maximum, and average similarity of the candidate structures to the target during the inversion process. Similarity S$_{0.1}$ is calculated as the radial basis function similarity with a width of $\gamma=0.1$. The similarity threshold to stop the algorithm is set to S$_{\mathrm{thresh}} = 0.99$.
We first discuss the top panel, which shows the genetic evolution of the descriptor inversion for Ag$_2$Se$_4$Y$_2$. To recall, stages 1 and 4 are exploration stages, stages 2 and 5 are optimization stages, and stages 3 and 6 are exploitation stages. In the first stage, the maximum similarity increases gradually to S$_{0.1}\approx0.06$, while the minimal and mean similarities remain close to 0. In the subsequent optimization stage, both the maximal and average similarity increase, while the minimal value does not. This indicates a diverse population, meaning that not only the best candidates are maintained in the population. In the third stage, the mean similarity instantly jumps to a higher value, as the algorithm favors the best structures in exploitation stages. However, the maximum similarity starts to saturate quickly at S$_{0.1}\approx0.20$, suggesting that the search got confined to a local minimum. To escape local minima and regain diversity, if no inversion of the descriptor was found in an exploitation stage, the population is reset to its state after the previous optimization stage and randomly perturbed. This leads to the initial dip of similarity in the beginning of stage 4. Afterwards, the maximum similarity rises sharply, reaching S$_{0.1}\geq$ S$_{\mathrm{tresh}}$ in stage 6 and triggering the algorithm to stop, since the final structure has a descriptor that is sufficiently similar / close to identical to the target. Visually (see Fig.~\ref{fig:structure-development}), the crystal structure is nearly indistinguishable from that of the target.

The Si$_2$O$_4$ case is shown in the middle panel. Here, the maximal similarity rises to S$_{0.1}\approx0.78$ already in the first stage, while the mean reaches S$_{0.1}\approx0.20$. In the next stage, the maximal similarity continues to rise, while the mean stays almost constant after an initial increase. In the following exploitation stage, the maximal similarity rises continuously, but does not reach the required breaking condition. The mean similarity stagnates during this stage. After this non-successful exploitation phase, the population is reset, as reflected by the drop of mean and maximal similarities. In the subsequent stages, the similarity rises continuously until the breaking condition is met in the following exploitation stage. The high values of the maximal similarity during the whole process can be also understood by comparing the structures with the target, all shown in Fig.~\ref{fig:structure-development}.

For the third example, FeCl$_2$, which is shown in the bottom panel, the maximal similarity swiftly increases S$_{0.1}\approx0.97$ already in the first stage. Unit cells and atomic arrangements already closely resemble the corresponding target structure, and the goal is met after only three stages. Interestingly, even the minimal similarity gets close to the maximal value, meaning that all structures of the population are very similar to the target structure. This state is kept during the whole second stage. In the third stage, the breaking condition is met quickly. It is likely that changes to the structure introduced by the crossovers and mutations in the second stage are too strong to reach the required maximal similarity, because the atoms are moved away from their optimal positions. Only much more conservative GA parameters of the exploitation phase are suitable to optimize the structure to its final shape.

\subsection{Generalization and robustness}

\begin{figure}
	\centering
	\includegraphics[width=\textwidth]{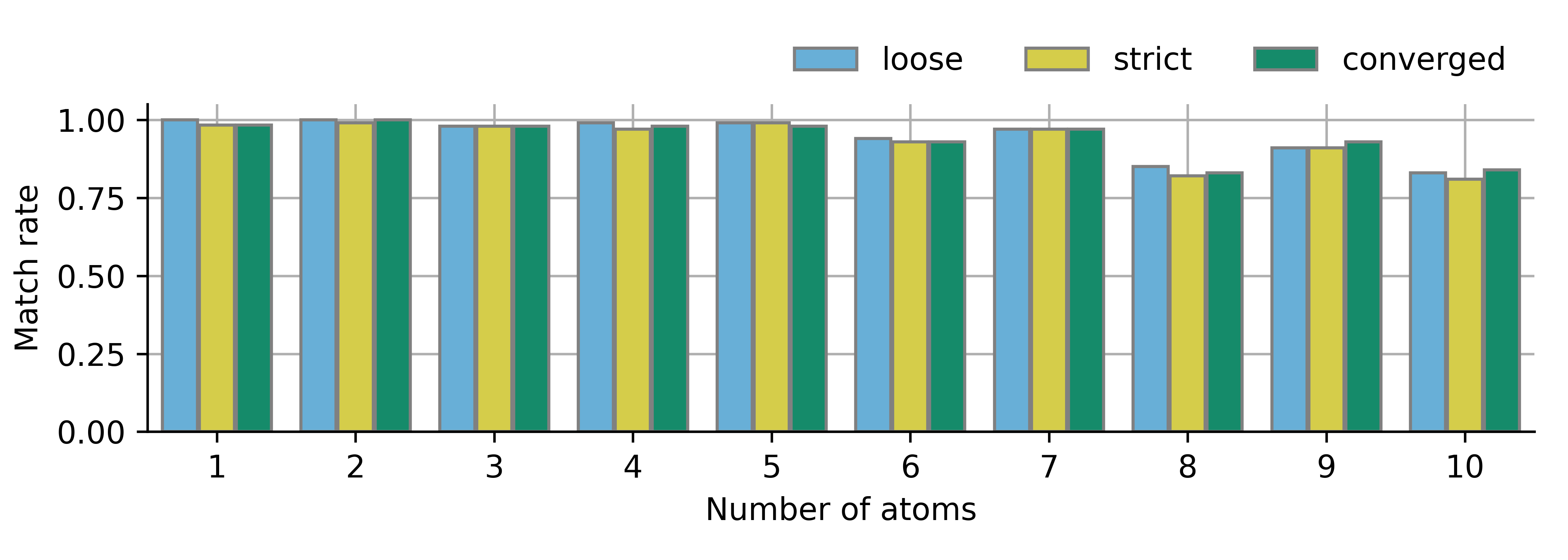}
	\caption{Match rates of inverted structures to their target with respect to the number of atoms in the unit cell (1-10 atoms). The blue and yellow bars show loose and strict matching with the \texttt{StructureMatcher} method \cite{Python_Material_Ong_S_2013}, respectively. Green bars show converged runs, \ie those reaching the breaking condition.}
	\label{fig:stats-multi-run}
\end{figure}

We verify the generality of our approach by applying the algorithm to different compounds containing 1 to 10 atoms per unit cell. For each number of atoms, we randomly select 100 structures from the mp-20 dataset~\cite{MP20-Github-Zugner}, with the exception of one-atomic structures were only 58 are available. This gives a total of 958 examples. For each of these structures, we generate a SOAP descriptor as the target. We run the optimization on a high-performance computing (HPC) system, using 64 CPUs on one compute node, resulting in a total runtime of 48 hours. Of the 958 runs, 901 reach the breaking condition, producing structures with a similarity of S$_{0.1} \geq$ S$_{\mathrm{thresh}} = 0.99$. We validate the inverted structures automatically by comparing them to the target structures using the \texttt{StructureMatcher} method implemented in the Python package \texttt{pymatgen}~\cite{Python_Material_Ong_S_2013}. To make our results better comparable to other efforts in the literature, we adopt two different matching criteria, "loose" and "strict", from Ref.~\cite{An_invertible_Xiao_2023}. A total of 905 structures (94.4\%) match their target structure with the loose criterion and 895 (93.3\%) with the strict criterion. A more detailed view on these results is presented in Fig.~\ref{fig:stats-multi-run}. The histogram shows the match rates as a function of system size, comparing the loose and strict settings of the \texttt{StructureMatcher} to the convergence criterion of our algorithm. It can be seen that for systems with up to 5 atoms, almost all structures are inverted successfully. After this point, the success rates decreases slowly as the system size increases, and reaches sightly more than 0.80 for compounds with 10 atoms in the unit cell. Curiously, for structures with even numbers of atoms (6, 8, 10), the match rate declines faster than for odd-numbered systems. For structures with up to 8 atoms, the convergence rates produced by our algorithm are nearly equal to the strict matching rates. This indicates that the current convergence condition is sufficient for structures with few atoms. For bigger structures (9-10 atoms) the convergence rate is slightly higher than even the rate with the loose condition. This shows that some structures reach the breaking condition even though the target structure is not found, \ie the result does not satisfy the criteria of the \texttt{StructureMatcher}. This can be counteracted by using a stricter threshold in our algorithm, since a higher descriptor similarity is expected to correlate with a higher structural similarity. We did not apply this in this example for the sake of demonstration.

\begin{figure}
	\includegraphics[width=\textwidth]{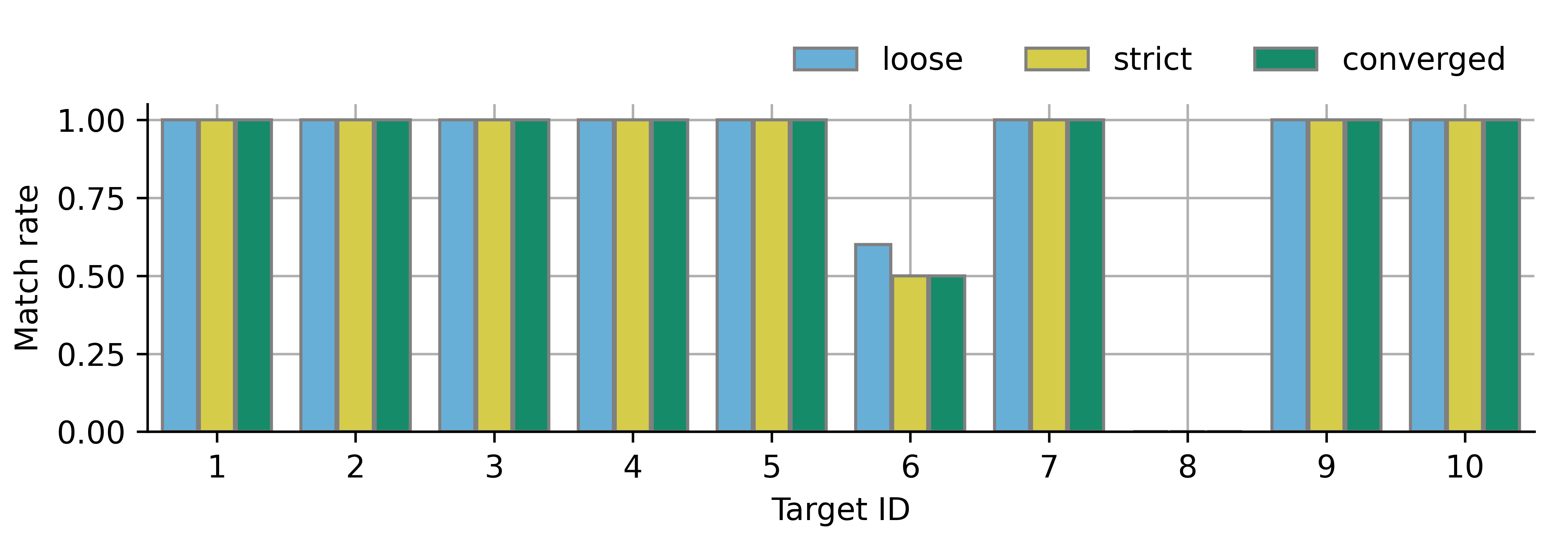}
	\caption{Match rates for 10 repeated inversion runs for 10 different target structures. The blue and yellow bars show loose and strict matching with the \texttt{StructureMatcher} method~\cite{Python_Material_Ong_S_2013}, respectively. Green bars show converged runs, \ie they reached the breaking condition.}
	\label{fig:robustness-analysis}
\end{figure}

We furthermore verify the robustness of our algorithm against randomness. We randomly select 10 materials (with up to 10 atoms) from the mp-20 dataset~\cite{MP20-Github-Zugner}, compute their descriptors, and repeat the structure inversion 10 times using different random seeds for the pseudo random number generator. Figure \ref{fig:robustness-analysis} presents the match rates of these 100 runs. Overall, 86\% of runs are successful (loose matching), however, with varying success rates: 8 out of 10 structures can be inverted with 100\% success rate. Yet, for the structure with ID 6, only 60\% success rate is achieved. Here, the mean similarity to the target descriptor is $\overline{S} = 0.8$, and the different results show a significant spread for individual runs. For the structure with ID 8, not a single inversion is successful. The runs reach a mean similarity of $\overline{S} = 0.2$, and most of them fail with identical results. A possible explanation is that some local extrema in the search space are hard to escape for the algorithm in its current configuration.

\subsection{Implementation}

To implement the inversion framework, we developed the open-source Python package \texttt{FUCrIMODo} (\textit{Find Unknown Crystals by Inversion of Machine-learning Optimized Descriptors}). The package provides a modular, extensible architecture that can be used both as a standalone program and as a Python library. While the present study focuses on the global SOAP descriptor, \texttt{FUCrIMODo} can be adapted to use any type of descriptor. The configuration of the algorithm, \eg the specific selection of parameters for genetic operations and the number of GA iterations per stage, which is used for the present work, has been found empirically. To be able to update the configuration efficiently, \texttt{FUCrIMODo} records different data during runtime, which can be visualized automatically using the analysis module of the software. To increase the reproducibility and data handling of all results, \texttt{FUCrIMODo} includes an additional package called \texttt{fucrimodo\_lab}, which is a structured framework for defining, organizing, analyzing, and storing runs. Due to high demand on computational power typical for genetic algorithms, the program provides parallelization optimized for execution on computer clusters.

\section{Discussion}\label{sec:Discussion}

This work introduces a novel method to invert descriptors for atomic structures, as demonstrated on the example of the global SOAP descriptor for many compounds of various structure types. The relationship between this descriptor and the atomic structure is highly nonlinear, caused by the averaging nature and the invariance of the descriptor to rotations and translations of the input structure. This makes the process of descriptor inversion highly non-trivial. Despite this challenge, our method successfully inverted the descriptors of a wide range of target structures consisting of up to 10 atoms, and we expect it to work for structures with more atoms in the unit cell. Summarizing, our approach has been showcased to reconstruct 94.4\% of structures from their SOAP descriptor, demonstrating its capability to perform inversion of descriptors that are not invertible analytically. A current limitation of the method is the degrading success with the structure size. It can be understood by the averaging process when constructing the global SOAP descriptor, which leads to an increasing loss of information for bigger structures.

Our approach is independent of the choice of the descriptor. However, the SOAP descriptor has specific properties that can be exploited to optimize the structure selection process, as described in Sections~\ref{sec:methods.descriptors} and \ref{sec:methods.fitness}. We expect similar performance for all types of continuous descriptors, \ie those, for which small changes to the structure lead to small changes in the descriptor~\cite{DScribe_Librar_Himane_2020}. The modular design of the \texttt{FUCrIMODo} code allows to adopt different descriptors, even though some parts of the code, such as the fitness function, are optimized towards SOAP and would require additional implementation, \eg in a \texttt{fucrimodo\_lab}.

A robustness test shows that for most target descriptors results do not change when using different random number generators. Just in one case, only 60\% of performed runs were successful within the given time frame. Repeated tests with increased runtime did not resolve the problem, which suggests that the configuration is not optimal for this structure. This can be addressed by using  \texttt{fucrimodo\_lab} to try different parameter sets and observing the success rate with the analysis module of \texttt{FUCrIMODo}. Being based on GA, a stochastic method, our approach shows some starting-point dependence, which can affect the total runtime of the algorithm. While some runs terminate after a few minutes, others need multiple hours. In the general case, where the number of atoms of the target is unknown, the algorithm can be run repeatedly, assuming different numbers of atoms, until a sufficiently high similarity to the target is achieved.

During the course of the project, different parametrizations of the global SOAP descriptor implemented in \texttt{DScribe} have been used (see also Section~\ref{sec:Methods}). Using the 'inner' averaging method, we encountered cases where the similarity score between two descriptors was very high, yet the structures were visibly different. This means that the descriptor obtained with this method is not unique or not continuous. When using the 'outer' method, this problem did not appear. This highlights another possible use case for our approach: Inverting descriptors can be used to test the properties of a descriptor, because the GA can uncover edge cases in which the descriptor performs unexpectedly.

Being able to invert a descriptor is not yet the complete answer to model inversion. Yet, it is an important step in this direction, as it allows to decouple the model, which is based on the descriptor, from the system.

\section{Methods}\label{sec:Methods}

\subsection{Descriptors} \label{sec:methods.descriptors}

The Smooth Overlap of Atomic Positions (SOAP) descriptor is a widely used representation of atomic structures in machine learning. Here, it serves as a prototypical descriptor for demonstrating our inversion method. Like many structural descriptors, it is designed to be invariant under rotation, translation, and permutation of identical atoms. It represents the local environment of each atom by a spatial density of all neighboring atoms that fall within a spherical cutoff radius. The density at each point in space is given by the superposition of Gaussians centered at the positions of the neighboring atoms. It is subsequently expanded in an orthonormal basis consisting of spherical harmonics and radial functions. This basis is used to construct the coefficients of power spectra, which themselves constitute the descriptor~\cite{On_representing_Bartok_2013, DScribe_Librar_Himane_2020}. To capture chemical identity, the expansion -- and hence the descriptor -- is computed separately for each elemental species and for all cross-species pairs present in the compound. The full SOAP descriptor of a local environment is expressed as a vector that is formed by concatenating the power spectrum components of all single-species and cross-species pairs. We obtain a global descriptor by computing the average of the SOAP vectors at the position of each atom. The descriptors are calculated using the Python package \texttt{DScribe} \cite{DScribe_Librar_Himane_2020, Updates_to_the_Laakso_2023} version 2.1.1.

The parameters are chosen to allow for the representation of a broad range of structures while keeping computational cost low \cite{DScribe_Librar_Himane_2020, Leveraging_gene_Barnar_2023}. We set the cutoff radius to $r_{\text{cut}}=15.0$~\AA, the radial basis size to $n_{\text{max}}=8$, the degree of spherical harmonics to $l_{\text{max}}=6$, and the Gaussian width $\sigma=0.5$~\AA, and we enable periodic boundary conditions for the descriptor generation. For the generation of the global descriptors, we chose the \textit{outer} averaging method. To minimize dimensionality, we include only those species that are present in the target structure. This eliminates zero-valued feature groups.

\subsection{Details of the algorithm}

\subsubsection{Structural sampling}

The initial population is generated from random sampling using the Python package \texttt{pyxtal}~\cite{PyXtal_A_Pytho_Freder_2021}. This sampling requires two ingredients, \ie the number of atoms, which is an input parameter of our multi-stage GA algorithm, and the chemical elements, which are inferred from the parameterization of the SOAP descriptor. The ratios of the elements are approximated by the summed-up magnitude of each species-specific block of the target SOAP features. These ratios only serve as a guideline for the initial sampling, and the composition can change during the later stages of the GA. Symmetry constraints based on the number of atoms are enforced by using the \texttt{symmetry} module of \texttt{pyxtal}. To form the initial population, the algorithm generates 5000 structures. Structures, where the distance between neighboring atoms is lower than the sum of their covalent radii, or where any lattice constant is bigger than 100 \AA, are eliminated. These constraints ensure that the generated structures follow basic physical rules and are as small as possible.

\subsubsection{Similarity measure}

To evaluate how similar the descriptor $\mathbf{D}(C)$ of a candidate structure $C$ is to the descriptor $\mathbf{D}(T)$ of the (unknown) target structure $T$, we choose the following similarity metric:
\begin{equation}
	\label{eq:rbf-similarity}
	S_{\gamma}(\mathbf{D}(T), \mathbf{D}(C)) = \exp\left(-\gamma \cdot ||\mathbf{D}(T) - \mathbf{D}(C)||^2\right),
\end{equation}
where $||\cdot||$ is the conventional $l_2$ norm and $\gamma$ is a parameter that controls the sensitivity of the metric. For the same pair of non-identical descriptors, a larger (smaller) value of $\gamma$ corresponds to smaller (larger) similarity. In all cases, we use $\gamma = 0.1$ to make results comparable.

\subsubsection{Fitness function} \label{sec:methods.fitness}

During GA stages, candidate structures are evaluated by so-called fitness functions $f$, allowing to consistently compare and distinguish between more and less optimal structures in the population. The fitness function can be chosen arbitrarily, however, the success of the GA is strongly dependent on this choice. Here, we use a combined fitness function that utilizes two different mechanisms to evaluate the fitness $f(C)$ of a candidate C. The first contribution is the similarity of the descriptor of the candidate structure to that of the target. This is based on the similarity measure defined above, leading to:
\begin{equation}
	f_\gamma^\mathrm{T}(C) = S_{\gamma}(\mathbf{D}(T), \mathbf{D}(C)) .
\end{equation}
Additionally, similarity is evaluated with species-specific fitness functions $f_{\gamma}^\mathrm{T, A, B}(C)$. These are calculated from the partial power-spectrum components of the SOAP descriptor, associated with atomic species pairs A and B.

The second fitness type measures the minimal distances between atoms in the unit cell. It is calculated as
\begin{equation}
	f^\mathrm{d}(C) = \exp\left[-\sum_{i=1}^{N} \sum_{j=i+1}^{N} \max\left(0, \frac{d^{\min}_{ij} - d_{ij}}{d^{\min}_{ij}}\right)\right],
\end{equation}
where $d_{ij}$ is the distance between atoms $i$ and $j$, and $d^{\min}_{ij}$ is the minimal allowed distance based on the ratio of covalent radii of the atoms. For different stages, different sets of fitness functions are used, however, $f^\mathrm{d}(C)$ and $f_{0.1}^\mathrm{T}$ are always used. During exploration and optimization, we additionally use $f_{1}^\mathrm{T}$ and $f_{0.01}^\mathrm{T}$. Further, we use species-specific fitness functions defined for all possible tuples (A, B) of atomic species. During optimization, we set $\gamma = 0.1$, during the first exploration $\gamma = 0.001$, and in all further exploration stages $\gamma = 0.5$.

\subsubsection{Population selection}

All of the GA stages are based on the ($\mu$+$\lambda$)-evolution strategy \cite{Evolution_strat_Beyer_2002}. In each generation of the GA, the genetic operators generate $\lambda$ new structures by modifying high-scoring members of the population, called \textit{parents}, with stage-specific crossovers and mutations. After combining these new individuals with the parent population, $\mu$ structures are selected to form the new population. This process requires selecting both parents as well as the members of the new population. For both selection processes, we use different approaches. To select parents, we use tournament selection implemented in the Python package \texttt{deap}~\cite{DEAP_JMLR2012}, in which $\lambda$ subsets of $n_T$ individuals are sampled randomly from the population, and the best-performing structure of each subset is selected. The tournament size n$_{T}$ strongly affects the diversity of the population. For large values of $n_{T}$, the resulting population will consist mostly of high scoring individuals. For smaller values, lower scoring individuals can also be selected for the new population. The standard tournament selection implemented in \texttt{deap} evaluates individuals based on a single fitness value, for which we use $f_{0.1}^\mathrm{T}$. We set $\lambda = 100$ and $n_{T} = 3$ in the optimization phase, and  $\lambda = 80$ and $n_{T} = 5$ in the exploitation phase. In the exploration phase, we set $\lambda = 250$ and use a special variant of the tournament selection (\texttt{deap.tools.selTournamentDCD}), which explicitly considers individual contributions to the fitness score instead of a single fitness value.

Once the parents have been selected, the genetic operators are applied to generate the offspring. In each stage, the specific mutation and crossover operations are drawn randomly from the pool of stage-specific operators, so that each offspring is produced by a randomly chosen combination of operations. Whether a given operator is applied to a selected individual is controlled by a mutation probability  P$_\mathrm{m}$ or crossover probability P$_\mathrm{c}$. We use P$_\mathrm{m}$ = 0.2 and P$_\mathrm{c}$ = 0.5 in the optimization stage, P$_\mathrm{m}$ = 0.2 and  P$_\mathrm{c}$ = 0.8 in the exploration stage, and P$_\mathrm{m}$ = 0.5 and  P$_\mathrm{c}$ = 0.5 in the exploitation stage. Applying the operators probabilistically, allows offspring to be created through mutation or crossover alone, rather than requiring both simultaneously. This increases the variety of structures that can be generated.

After applying genetic operators, we select $\mu$ = 500 (exploration and optimization stage) or $\mu$ = 100 (localization stage) individuals from the combined population using NSGA-II selection \cite{A_fast_and_elit_Deb_K_2002} to preserve the highest-scoring individuals while maintaining diversity. NSGA-II selection first chooses all individuals with the highest values for all objectives and then fills up the remaining population based on two additional metrics, which are rank and crowding distance. The rank of individuals is determined by comparing the different fitness values. Crowding distance quantifies the isolation of an individual from its immediate neighbors by considering individual components of the fitness score.

\subsubsection{Validation of inverted structures}

We validate the success rate of the inversion process by comparing the structure with the highest similarity obtained by the algorithm to the target structure. As mentioned already, invariance of the atomic structure with respect to translations and rotations of the lattice requires the usage of additional metrics. We therefore make use of the \texttt{StructureMatcher} algorithm as implemented in \texttt{pymatgen}~\cite{Python_Material_Ong_S_2013}. It performs multiple transformations to match both structures within predefined tolerances. Two tolerance settings are used to make results comparable to previous works related to descriptor inversion~\cite{An_invertible_Xiao_2023}: \textit{loose} (\textit{strict}) settings use fractional length, site, and angle tolerance of 0.3 (0.2), 0.5 (0.3), and 10$^{\circ}$ (5$^{\circ}$), respectively.

\subsection*{Data Availability}

All data that was produced for obtaining the results can be found at Ref.~\cite{data}.

\subsection*{Code availability}

The code is available on GitHub (\url{https://github.com/OHANAN1/fucrimodo}) under the Apache 2.0 license. Installation instructions and documentation can be found at \url{https://fucrimodo.readthedocs.io/en/latest/index.html}. The scripts that produce the main results of this work are automatically installed together with the program and can be run with the instructions in the documentation.

\section{Acknowledgements}

This work received funding from the German Research Foundation (DFG) through the CRC 1404 (FONDA), project 414984028 and the NFDI consortium FAIRmat, project 460197019. We further acknowledge compute time provided by the Humboldt-Universität zu Berlin via HPC@HU.

\subsection*{Author Contributions}

L.B.: Software, Methodology, Investigation, Writing - original draft.
M.K.: Conceptualization, Supervision, Validation, Writing - review \& editing.
C.D.: Conceptualization, Supervision, Writing - review \& editing.

\section{Competing Interests}

The authors declare no competing interests.

\end{document}